\documentclass[prd,eqsecnum,apsi]{revtex4}

\usepackage{mathrsfs}
\usepackage{graphicx}
\usepackage{latexsym}
\usepackage{amsmath}
\usepackage{amssymb}
\begin{document}
\title{\Large \bf Antisymmetric tensor fields in a generalized Randall-Sundrum scenario}
\author{Ashmita Das \footnote{Electronic address: {\em tpad@iacs.res.in}}
${}^{}$and Soumitra SenGupta \footnote{Electronic address : {\em tpssg@iacs.res.in}
}${}^{}$}
\affiliation{Department of Theoretical Physics\\
Indian Association for the Cultivation of Science,\\
Kolkata - 700 032, India}

\begin{abstract}
Bulk antisymmetric tensor fields of different ranks have been studied in the context of a generalized Randall-Sundrum 
model with a non-vanishing induced cosmological constant on the visible brane. It is shown that instead of the usual 
exponential suppression
of the couplings of the zero modes of these bulk fields with the brane fermions in the original Randal-Sundrum model ,
here the couplings are proportional
to the brane cosmological constant. Thus in an era of large cosmological constant these fields have significant role
in physical phenomena because of their enhanced couplings with the visible brane fermions. 
\end{abstract}
\maketitle
Large hierarchy of mass scales between the Planck and the Tev scales results into the well-known fine tuning problem
in connection with Higgs the only scalar particle in the standard model. 
It has been shown that due to large radiative corrections the  Higgs mass can not be confined within Tev unless some unnatural tuning
is done order by order in the perturbation theory. The two most successful  efforts to resolve this crisis are 
supersymmetry\cite{Martin,Dress} and/or extra dimensional generalization of standard model both of which  lead us to the Physics
beyond standard model ( BSM )\cite{ADD,RS}. Among various extra dimensional models, the warped geometry 
model proposed by Randall and Sundrum\cite{RS} has drawn special attention
for the following reasons : (1) It resolves the hierarchy problem without introducing any other hierarchial scale in the theory,
(2) The modulus of the extra dimensional model can be stabilized \cite{GW1}, (3) It provides interesting new phenomenology which can be
tested in the Tev scale collider experiments say in LHC and  (4) A warped solution , though not exactly same as RS model, can be found
from string theory which as a fundamental theory predicts inevitable existence of extra dimensions \cite{Green}.\\
Randall-Sundrum scenario \cite{RS} which is defined on a 5-dimensional anti de-Sitter space-time with one spatial
direction orbifolded on $S^1/Z_2$  has the following features : \\
\begin{itemize}
\item Two 3-branes namely hidden/Planck brane and visible/standard model brane are located at the two orbifold fixed points.  
\item The effective  cosmological constant induced on the hidden and visible brane are zero i.e these are flat branes.\\
\item The brane tension of the standard model/visible brane is negative.\\
\item Without introducing any extra scale, other than the Planck scale, in the theory 
one can choose the brane separation modulus $r_c$ to have a value $M_P^{-1}$ such that the desired warping can be obtained between the
two branes from Planck scale to Tev scale.\\  
\item The modulus can be stabilized to the above chosen value 
by introducing scalar in the bulk \cite{GW1} without any further fine tuning.\\ 
 \end{itemize}
In this model it is assumed that all the standard model fields are confined on the visible brane while the gravity
propagates in the bulk. The main motivation behind this assumption has its root in string theory where the SM fields are
open string modes whose end points are fixed on the brane while gravity being a closed string mode can reside in the bulk.
Following this argument all the antisymmetric tensor string modes of various ranks are also expected to
propagate in the bulk. Despite having similar coupling with brane matter just as graviton mode none of these antisymmetric tensor modes
so far has been detected through any experimental signature. As an explanation of their invisibility
it has been shown that in a warped geometry model all the antisymmetric modes of two or higher ranks 
are suppressed by successive higher powers of the exponential warp factor on the visible brane \cite{Ssg1,Ssg2,Lebedev}.\\

Various phenomenological as well as cosmological implications of RS model have been
discussed in several works\cite{Grossman,Agashe,Davoudiasl2,Ssg4,Ssg5,Ssg6,Ssg7,K.Dasgupta,J.M.hoff,Gwyn,Nelson,Rubakov}.
Meanwhile RS model has been generalized \cite{Ssg3} such that the visible 3-brane can either be de-Siter or anti-de-Sitter with positive
or negative induced cosmological constant. Such models not only can resolve the gauge hierarchy problem but also 
may render stability to the visible
brane which now can be endowed with positive tension. This work was motivated by the facts that the zero 
cosmological constant of the visible 3-brane is not consistent with the observed 
small value of the cosmological constant of our Universe and negative tension
branes are intrinsically unstable. Moreover the possibility of a de-Sitter universe by antisymmetric tensor flux compactification
has been shown in the context of string inspired supergravity models \cite{Kachru}. Such models with flux and branes are known to have a generic warped 
geometric structure. These lead us to explore the correlation between a non-vanishing 3-brane cosmological constant 
and the antisymmetric tensor fields on the brane. From a different viewpoint the connection between the cosmological constant and 
background space-time torsion has been studied \cite{Sabbata}. The third rank
field strength corresponding to the second rank anti-symmetric closed string mode namely the Kalb-Ramond field can be identified
with space-time torsion \cite{Hammond}. This field has been shown to have a highly suppressed coupling to the standard model fields in a
warped geometry model on a flat 3-brane. It is therefore important to explore whether such suppression leading to
an illusion of a vanishing torsion persists even when the space-time has non-vanishing cosmological constant.\\   

We first briefly outline the generalized RS model below.\\

The  warp factor in such a model is obtained by extremising the following the action :
\begin{equation}
S = \int d^5x \sqrt{-G} ( M^3 {\cal R} - \Lambda) + \int d^4x
\sqrt{-g_i} {\cal V}_i\label{eq1}
\end{equation}
where $\Lambda$ is the bulk cosmological constant, ${\cal R}$ is
the bulk ($5$-dimensional) Ricci scalar and ${\cal V}_i$ is the
tension of the $i^{th}$ brane ($i = hid(vis)$ for the hidden
(visible) brane). It is shown that a  warped geometry results from a constant 
curvature brane space-time, as opposed to a flat 3-brane space-time. The generalized ansatz for the warped metric is given by, 
\begin{equation}
ds^2 = e^{- 2 A(y)} g_{\mu\nu} dx^{\mu} dx^{\nu} +r^2 dy^2 \label{eq2}
\end{equation}
where  $r$ corresponds to the modulus associated with the extra dimension   and $\mu , \nu$ stands for brane coordinate 
indices. As in the original RS model, the scalar mass warping is achieved through the warp factor
$e^{-A(k r \pi)}=\frac{m}{m_0}=10^{-n}$ where $k = \sqrt{- \frac{\Lambda}{12 M^3}}$ $\sim$  Planck 
Mass with the bulk cosmological constant $\Lambda$ is chosen to be negative. `$n$' the 
warp factor index must be set to $16$ to achieve the desired warping 
and the magnitude of the induced cosmological 
constant on the brane in this case is non-vanishing in general and is given by =$10^{-N}$(in planckian units).  
A careful analysis reveals that for negative brane cosmological constant 
N has minimum value given by $N_{min}=2 n$ leading to an upper bound on the magnitude of the cosmological constant 
while there is no such upper bound for the induced positive cosmological constant on the brane.
Furthermore for the induced brane cosmological constant, $\Omega > 0$ and $\Omega < 0$, the brane metric $g_{\mu\nu}$ 
corresponds to some de-sitter or anti de-Sitter space-time say for example dS-Schwarzschild and AdS-Schwarzschild 
space-times respectively \cite{Alwis,Verlinde,Navarro,C.Csaki}.

For AdS bulk i.e.  $\Lambda<0$, considering the regime for which the induced cosmological constant $\Omega$ on the visible brane is 
negative if one redefines $\omega^2 \equiv -\Omega/3k^2 \geq 0$, then the following solution for the warp factor
is obtained :
\begin{equation}
e^{-A(y)} = \omega \cosh\left(\ln \frac {\omega} c_1 + ky \right)\label{wfads}
\end{equation}
where $c_1 = 1 + \sqrt{1 - \omega^2}$ for the warp factor normalized to unity at $y = 0$.\\        

Similarly when the induced brane cosmological constant is 
positive i.e the 3-brane is de-Sitter with $\Omega>0$, the warp factor turns out to be, 
\begin{eqnarray}
 e^{-A(\phi)} = \omega  {\rm sinh}|({\rm ln}\frac{c_2}{\omega}-kr_{c}\phi)|\label{wfds}\\
 {\rm where} , c_{2} = 1+\sqrt{1+\omega^{2}} , \omega^{2}=\frac{\Omega}{3k^{2}}\nonumber
\end{eqnarray} 
In order to resolve the gauge hierarchy problem the warp factor $ e^{-A(\phi)}$ must be equal to $10^{-16}$ at $\phi = \pi$ and 
this implies that for both anti de-Sitter and de-Sitter branes the values of $kr$ depend on the values of the cosmological
constant $\omega^2$. The RS solution namely $kr \sim 11.5$ for brane cosmological constant $\omega^2 =0$ is just one solution in the
plot of solutions in figure 1.
\begin{figure}[h] 
\includegraphics[width=3.330in,height=2.20in]{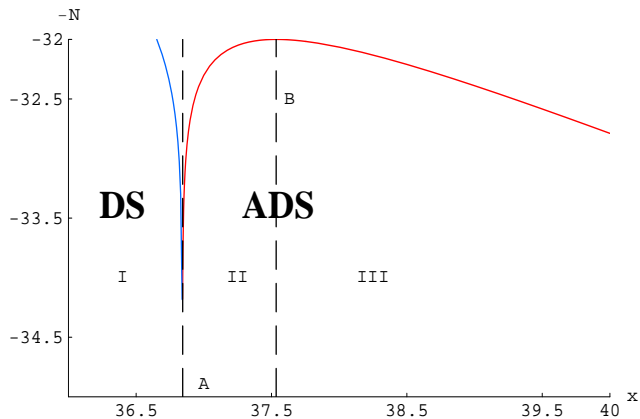}
\caption{Graph of $N$ versus $x = k r \pi =36-40$, for $n=16$ and for both
positive and negative brane cosmological constant. The curve in
region-I corresponds to positive cosmological constant on the brane,
whereas the curve in regions-II \& III represents negative
cosmological constant on the brane. } \label{ADSDS}
\end{figure}
Apart from the bulk graviton various other bulk fields like scalars, gauge and fermions fields have been considered in different work to
obtain their massless as well as massive KK towers on the visible 
brane \cite{Davoudiasl1,GW2}. However in the context of string theory where many higher rank
anti-symmetric tensor fields appear as closed string modes \cite{Duff}, the study of bulk fields have been 
widened to include these fields also. 
One such field namely the second rank antisymmetric tensor field ( called Kalb-Ramond field )\cite{KR} with third rank antisymmetric field 
strength can be viewed upon as the torsion field in the background space-time. The apparent torsion-free universe implies that such field,
if exists, must be heavily suppressed on the visible brane. It was then shown that a Randall-Sundrum warped geometry model can
indeed explain such a suppression of this field on the visible brane compared to the graviton through the large exponential 
warping which appears in the space-time metric. Inspired by this result the authors
of \cite{Ssg1} then extended this calculation for even higher rank antisymmetric  tensor fields which inevitably appear in string-based
models. It turned out that all such higher rank fields are even more suppressed by higher powers of the 
warp factors \cite{Ssg2} and Randall-Sundrum
model thus can explain the apparent invisibility of all these antisymmetric tensor fields in our universe.\\  

Here, we propose to re-examine this feature in the context of  
generalized RS model described earlier. Our prime concern is to find out possible modifications in the projections of the 
antisymmetric tensor fields on the (3+1)dimensional brane due to the inclusion of 
cosmological constant induced on the 3-brane.\\
We organize our work as follows :\\
\begin{itemize}
\item To determine the massless and the massive KK-modes of the two-form KR field and also of the higher rank antisymmetric
tensor field for anti-deSitter visible brane\\
\item To repeat the same calculation for the deSitter visible brane.\\ 
\item We also determine the coupling of KR field as well as the higher rank 
antisymmetric tensor fields to the fermion fields localized on the visible brane for both de-Sitter and anti de-sitter branes
to look for  their possinle presence through interactions with brane fermions. 
\end{itemize}

We consider space-time with torsion in a generalized Randall-Sundrum scenario,that is with a 
cosmological constant $\Omega$ induced on the visible brane. We recall from our previous discussions that the warp metric is,
\begin{eqnarray}
 ds^2=e^{-2A(\phi)}g_{\mu\nu}dx^{\mu}dx^{\nu}+r^2_{c}d^2{\phi}\label{metricgrs}\\
{\rm where}, e^{-A(\phi)}=\omega {\rm cosh}({\rm ln}\frac{\omega}{c_1}+kr_{c}{\phi})\nonumber\\
{\rm and}, c_1=1+\sqrt{1-\omega^2} ,\omega^2= -\frac{\Omega}{3k^2}\nonumber
\end{eqnarray}
\section*{\Large\bf Modes of the Antisymmetric tensor fields in  anti de-Sitter 3-brane}
We begin with the second rank antisymmetric tensor field ( namely the Kalb-Ramond field) with a third rank
tensor field strength.
As mentioned earlier we can identify the background space-time torsion with the rank 3 antisymmetric field strength tensor $H_{MNL}$ 
corresponding to the second rank anti-symmetric Kalb-Ramond tensor field $B_{MN}$ which are related 
as  $H_{MNL}=\partial_{[M}B_{NL]}$. The 
KR gauge invariance allows us to set $B_{4\mu}=0$.\\
Using the explicit form of generalized RS metric and keeping $B_{4\mu}=0$ the action is given as,
\begin{eqnarray}
 S_{H}=\int d^4x\int d\phi r_{c}e^{2A(\phi)} [\eta^{\mu\alpha}\eta^{\nu\beta}\eta^{\lambda\gamma}H_{\mu\nu\lambda}H_{\alpha\beta\gamma}-\frac{3}{r^2_{c}}e^{-2A(\phi)}\eta^{\mu\alpha}\eta^{\nu\beta}B_{\mu\nu}\partial^2_{\phi}B_{\alpha\beta}]\label{actionkrads}
\end{eqnarray}
Using the Kaluza-Klein decomposition for the KR field:
\begin{eqnarray}
B_{\mu\nu}(x,\phi)=\sum^{\infty}_{n=0}B_{\mu\nu}^{n}(x)\frac{\chi^n(\phi)}{\sqrt{r_c}}\label{kk1}
\end{eqnarray}
the effective four dimensional action becomes
\begin{equation}
 S_{H}=\int d^4x\sum^{\infty}_{n=0}[\eta^{\mu\alpha}\eta^{\nu\beta}\eta^{\lambda\gamma}H^{n}_{\mu\nu\lambda}H^{n}_{\alpha\beta\gamma}+3m^{2}_{n}\eta^{\mu\alpha}\eta^{\nu\beta}B^{n}_{\mu\nu}B^{n}_{\alpha\beta}]\label{4daction1}
\end{equation}
provided
\begin{equation}
 -\frac{1}{r^2_{c}}\frac{d^{2}\chi^{n}}{d\phi^{2}}= m^{2}_{n}\chi^{n}e^{2A(\phi)}\label{dif1}
\end{equation}
along with the orthonormality condition
\begin{equation}
 \int e^{2A(\phi)}\chi^m(\phi)\chi^n(\phi)d\phi=\delta_{mn}\label{oc1}
\end{equation}
In terms of $z_{n}=\frac{m_{n}}{k}e^{A(\phi)}$ equation (\ref{dif1}) can be recast in the form
\begin{eqnarray}
&& z_{n}^2\frac{d^{2}\chi^n}{dz_{n}^{2}}+z_{n}\frac{d\chi^{n}}{dz_{n}}-\frac{a^2z_{n}^{3}}{(1-a^{2}z_{n}^{2})}\frac{d\chi^{n}}{dz_{n}}+\frac{z_{n}^{2}\chi^{n}}{(1-a^{2}z_{n}^{2})}=0\label{zn1adskr}
 \end{eqnarray}
where, $a^{2}=\frac{k^{2}\omega ^{2}}{m_{n}^{2}}\nonumber$

Keeping the leading order terms we get,
\begin{equation}
[z_{n}^{2}\frac{d^{2}}{dz_{n}^{2}}+z_{n}\frac{d}{dz_{n}}+z_{n}^{2}(1+a^{2}z_{n}^{2})]\chi_{n}=0\label{zn2adskr}
\end{equation}
The solution of the above equation can be written as,
\begin{equation}
 \chi^{n}(\phi) = \frac{1}{N_{n}}[J_{0}(z_{n})+\alpha_{n}Y_{0}(z_{n})+\omega^{2}\xi^{n}]\label{mmkrads1}
\end{equation}
From continuity condition at $\phi=0$ we obtain , $\alpha_{n} \simeq x_{n}e^{-A(\pi)}$
where $x_{n}=z_{n}(\pi)$. This implies $\alpha_{n}<<1$.
Also at $\phi=\pi$ we get, $J_{1}(x_{n})=\frac{\pi}{2}x_{n}e^{-A(\pi)}$.\\
The differential equation for $\xi^{n}$ now becomes,
\begin{eqnarray}
 \left[ z_{n}^{2}\frac{d^2}{dz_{n}^2}+z_{n}\frac{d}{dz_{n}}+z_{n}^{2}\right] \xi_{n}+\frac{k^{2}}{m_{n}^{2}}z_{n}^{4}J_{0}(z_{n})=0\label{zn3adskr}
\end{eqnarray}
To the leading order $J_{0}(z_{n})=\frac{1}{2}$ and the above differential equation becomes
\begin{eqnarray}
\left[ z_{n}^{2}\frac{d^2}{dz_{n}^2}+z_{n}\frac{d}{dz_{n}}+z_{n}^{2}\right] \xi^{n}+\frac{1}{2}\frac{k^{2}}{m_{n}^{2}}z_{n}^{4}=0\label{zn4adskr}
\end{eqnarray}
We therefore obtain the solution for $\xi^{n}$ as,
\begin{eqnarray}
 \xi^{n}&=&\frac{1}{4}\frac{k^{2}}{m_{n}^{2}}[2\pi z_{n}^2J_{0}(z_{n})Y_{0}(z_{n})-4\pi z_{n}J_{0}(z_{n})J_{1}(z_{n})Y_{0}(z_{n})-2\pi\nonumber\\
&&z_{n}^{2}J_{2}(z_{n})Y_{0}(z_{n})+\pi z_{n}^{3}J_{3}(z_{n})Y_{0}(z_{n})-4\pi z_{n}J_{0}(z_{n})Y_{1}(z_{n})\nonumber\\
&&+\pi z_{n}^{3}J_{0}(z_{n})Y_{1}(z_{n})+4\pi z_{n}J_{0}(z_{n})^{2}Y_{1}(z_{n})]\label{xin1}
\end{eqnarray}
Using the orthonormality condition (\ref{oc1}) and performing the numerical integration we find,
\begin{eqnarray}
 N_{n}=\frac{1}{\sqrt{kr_{c}}}\frac{k}{m_{n}}\left[0.712-\frac{k^{2}\omega^{2}}{m_{n}^{2}}0.026\right] ^{\frac{1}{2}}\label{Nn1}
\end{eqnarray}
The final solution for the massive modes turns out to be,
\begin{eqnarray}
 \chi_{n}(z_{n})=\sqrt{kr_{c}}\frac{m_{n}}{k}\left[ 0.712-\frac{k^{2}\omega^{2}}{m_{n}^{2}}0.026\right]^{-\frac{1}{2}} [J_{0}(z_{n})+\omega^{2}\xi^{n}]\label{fskrads}
\end{eqnarray}

We now turn our attention to massless mode. The differential equation for the massless mode is,
\begin{eqnarray}
 \frac{1}{r_{c}^{2}}\frac{d^{2}\chi^{n}}{d\phi^{2}}=0\label{mlmdif1}
\end{eqnarray}
The solution of the above equation is, 
\begin{eqnarray}
\chi^{0}(\phi)=c_{1}\phi+c_{2}\label{smlmkrads}
\end{eqnarray}
Applying the continuity condition we find $c_{1}=0$.\\
Hence, we get $\chi^{0}=c_{2}$. Now using orthonormality condition,
\begin{eqnarray}
 \int^{\pi}_{0} e^{2A(\phi)}c_{2}^{2}  d\phi=1\label{useoc1}
\end{eqnarray}
We obtain $c_{2}^{2}=2kr_{c}e^{-2kr_{c}\pi}\left[ 1+\frac{\omega^{2}}{4}(1+e^{2kr_{c}\pi})\right]$.
Plugging in the solution of $c_{2}$ we finally arrive at the  expression for $\chi^{0}$ as,
\begin{eqnarray}
 \chi^{0}=\sqrt{2}\sqrt{kr_{c}}e^{-kr_{c}\pi}\left[ 1+\frac{\omega^{2}}{4}(1+e^{2kr_{c}\pi})\right] ^{\frac{1}{2}}\label{mlms}
\end{eqnarray}
This the solution for the massless KR mode on the brane.\\

It may be observed that both the massless and the massive
KR mode depend on the induced brane cosmological constant $\omega^{2}$.
For $\omega = 0$ i.e when the warped geometry corresponds to RS model,  we retrieve the result that the KR field is heavily suppressed 
on the visible brane as obtained in \cite{Ssg1}. 

Let us now generalize the above analysis and consider the bulk antisymmetric tensor field of  higher rank i.e a rank-3 tensor, $X_{MNA}$, 
with the corresponding field strength tensor $Y_{MNAB}$.
\begin{eqnarray}
 S=\int d^{5}x \sqrt{-G} Y_{MNAB}Y^{MNAB}\label{actionyads}
\end{eqnarray}
G is the determinant of the 5 dimensional metric.In general one should be able to write down a rank-$(n+1)$ antisymmetric tensor 
field strength tensor as,
\begin{eqnarray}
 Y_{a_{1}a_{2}....a_{n+1}}=\partial_{[a_{n+1}} X_{a_{1}a_{2}....a_{n}]}\label{yads1}
\end{eqnarray}
Using the explicit form of the generalized RS metric and using the gauge fixing condition i.e $X_{\mu\nu y}=0$ one obtains,
\begin{eqnarray}
 S_{x}=\int d^{4}x\int d\phi[e^{4A(\phi)}\eta^{\mu\lambda}\eta^{\nu\rho}\eta^{\alpha\gamma}\eta^{\beta\delta}Y_{\mu\nu\alpha\beta}Y_{\lambda\rho\gamma\delta}+ 4 \frac{e^{2A(\phi)}}{r_{c}^{2}} \eta^{\mu\lambda} \eta^{\alpha\delta} \eta^{\nu\rho} \partial_{\phi}X_{\mu\nu\alpha} \partial_{\phi}X_{\lambda\rho\gamma}]\label{action5yads}
\end{eqnarray}
Considering the KK decomposition of the field X,
\begin{eqnarray}
 X_{\mu\nu\alpha}(x,\phi)= \sum_{n=0}^{\infty}X_{\mu\nu\alpha}^{n}(x) \frac{\chi^{n}(\phi)}{\sqrt{r_{c}}}\label{kkyads}
\end{eqnarray}
an effective action can be obtained for the projection $X_{\mu\nu\alpha}$ on the visible brain,
\begin{eqnarray}
 S_{x}=\int d^{4}x \sum_{n}[\eta^{\mu\lambda}\eta^{\nu\rho}\eta^{\alpha\gamma}\eta^{\beta\delta}Y_{\mu\nu\alpha\beta}^{n}Y_{\lambda\rho\gamma\delta}^{n}+ 4 m_{n}^{2}\eta^{\mu\lambda}\eta^{\nu\rho}\eta^{\alpha\delta} X_{\mu\nu\alpha}^{n}X_{\lambda\rho\gamma}^{n}]\label{action4yads}
\end{eqnarray}
where $m_{n}^{2}$ is defined through the relation,
\begin{eqnarray}
 -\frac{1}{r_{c}^{2}}\frac{d}{d\phi}(e^{2A(\phi)}\frac{d}{d\phi}\chi^{n})= m_{n}^{2}\chi^{n}e^{4A(\phi)}\label{difyads}
\end{eqnarray}
$\chi^{n}$ satisfies the orthonormality condition,
\begin{eqnarray}
&& \int e^{4A(\phi)}\chi^{m}(\phi)\chi^{n}(\phi) d\phi=\delta_{mn}\label{ocyads}
\end{eqnarray}
Introducing $f_{n}=e^{A(\phi)}\chi^{n}$ equation (\ref{difyads}) can be recast in the form,
\begin{eqnarray}
\left[ z_{n}^{2}\frac{d^{2}f_{n}}{dz_{n}^{2}}+z_{n}\frac{df_{n}}{dz_{n}}+\frac{f_{n}z_{n}^{2}}{(1-a^{2}z_{n}^{2})}-f_{n}+f_{n}\frac{a^{2}z_{n}^{2}}{(1-a^{2}z_{n}^{2})}-z_{n}^{3}a^{2}\frac{df_{n}}{dz_{n}}\right] = 0\label{znyads1}
\end{eqnarray}
where,  $z_{n}=\frac{m_{n}}{k}e^{A(\phi)}$ and   $a^{2}=\frac{k^{2}\omega^{2}}{m_{n}^{2}}$. 
Ignoring the last term in comparison to the term containing $z_{n}$
we find,
\begin{eqnarray}
\left[ z_{n}^{2}\frac{d^{2}}{dz_{n}^{2}}+z_{n}\frac{d}{dz_{n}}+\frac{z_{n}^{2}}{(1-a^{2}z_{n}^{2})}-1+\frac{a^{2}z_{n}^{2}}{(1-a^{2}z_{n}^{2})}\right] f_{n} =0\label{znyads2}
\end{eqnarray}
The solution of the above equation can be written as,
\begin{eqnarray}
&& f_{n}=\frac{1}{N_{n}}[J_{1}(z_{n})+\alpha_{n}Y_{1}(z_{n})+\omega^{2}\xi^{n}]\nonumber\\
&&\chi^{n}=e^{-A(\phi)} f_{n}=\frac{e^{-A(\phi)}}{N_{n}}\left[ J_{1}(z_{n})+\alpha_{n}Y_{1}(z_{n})+\omega^{2}\xi^{n}\right]\label{chinyads1}
\end{eqnarray}
We can reduce the equation (\ref{znyads2}) into the differential equation for $\xi^{n}$ as,
\begin{eqnarray}
 z_{n}^{2} \frac{d^{2}\xi^{n}}{dz_{n}^{2}} + z_{n} \frac{d\xi^{n}}{dz_{n}} + [z_{n}^{2}(1+a^{2})-1]\xi^{n}=0\label{znyads3}
\end{eqnarray}
The solution for $\xi^{n}$ turns out to be, 
\begin{eqnarray}
\xi^{n}&=&\alpha_{n}Y_{1}(\sqrt{1+a^{2}}z_{n})-J_{1}(-\sqrt{1+a^{2}}z_{n})\nonumber\\
\chi^{n} &=& \frac{e^{-A(\phi)}}{N_{n}}[J_{1}(z_{n})+\alpha_{n}Y_{1}(z_{n})+\omega^{2}\{\alpha_{n}Y_{1}(\sqrt{1+a^{2}}z_{n})\nonumber\\
&&-J_{1}(-\sqrt{1+a^{2}}z_{n})\}] \label{chinyads1}
\end{eqnarray}
The desired mass value of $M_{n}$ on the visible brane should be of the order of the 
Tev scale $(<<k)$.
Using the continuity condition at $\phi=0$ and noting that $e^{kr_{c}\pi}>>1$ , 
we find
\begin{eqnarray}
\alpha_{n}=-\frac{J_{2}\left[ \frac{m_{n}}{k}\left( 1-\frac{\omega^{2}}{4}\right) \right] + \frac{\omega^{2}m_{n}^{2}}{8k^{2}} }{Y_{2}\left[ \frac{m_{n}}{k}\left( 1-\frac{\omega^{2}}{4}\right) \right]-\frac{4k^{2}\omega^{2}}{\pi m_{n}^{2}}\frac{1}{\left( 1-\frac{\omega^{2}}{2}\left( 1-\frac{2k^{2}}{m_{n}^{2}}\right) \right)}}\label{alphanyads}
\end{eqnarray}
Estimating the order of $\alpha_{n}$, we get $\alpha_{n}<<1$.  We therefore can write,
\begin{eqnarray}
\chi^{n}= \frac{e^{-A(\phi)}}{N_{n}}\left[ J_{1}(z_{n})-\omega^{2}J_{1}(-\sqrt{1+a^{2}}z_{n})\right]\label{chinyads2}
\end{eqnarray}
Again from the continuity condition at $\phi=\pi$  we find,
\begin{eqnarray}
J_{2}(x_{n})+\omega^{2}J_{2}(\sqrt{1+a^{2}}x_{n}) = 0\label{besselyads}
\end{eqnarray}
Here, $x_{n}=z_{n}(\pi)=\frac{m_{n}}{k}e^{A(\pi)}$. Once again
from orthonormality condition (\ref{ocyads}) ,and performing the integration numerically , we can have the expression for $N_{n}$ as,
\begin{eqnarray}
&&N_{n}=\frac{k}{m_{n}}\frac{1}{\sqrt{kr_{c}}}\left[ 0.136+\omega^{2}0.347\right] ^{\frac{1}{2}}\label{Nnyads}\\
&&\chi^{n}=\frac{e^{-A(\phi)}m_{n}}{k}\sqrt{kr_{c}}\left[ 0.136+\omega^{2}0.347\right] ^{-\frac{1}{2}}\nonumber\\
&&\left[ J_{1}(z_{n})-J_{1}(-\sqrt{1+a^{2}}z_{n})\right]\label{chinyads3}
\end{eqnarray}
For massless mode the differential equation becomes,
\begin{eqnarray}
\frac{1}{r_{c}^{2}} \frac{d}{d\phi}\left(e^{2A(\phi)}\frac{d}{d\phi}\chi^{n} \right)=0\label{mlmdifyads}
\end{eqnarray}
Solving the above differential equation we derive the solution for massless mode;
\begin{eqnarray}
\chi^{0}= c_{1}\left[ \frac{\omega^{4}e^{2kr_{c}\phi}}{32kr_{c}}-\frac{e^{-2kr_{c}\phi}}{2kr_{c}}+\frac{\omega^{2}\phi}{2}\right] + c_{2}\label{chi01}
\end{eqnarray}
Applying continuity condition, we find ( just as in previous case ),
$c_{1}=0$  and  $\chi^{0}=c_{2}$.
This leads to  $\chi^{0}=c_{2}$. Finally applying orthonormality condition,
\begin{eqnarray}
 \int^{\pi}_{0} e^{4A(\phi)}c_{2}^{2}  d\phi=1\label{useocyads}
\end{eqnarray}
we arrive at $c_{2}^{2}=4kr_{c}e^{-4kr_{c}\pi}\left( 1+\frac{3}{4}\omega^{2}e^{2kr_{c}\pi}\right)$
which on substitution yields the final expression for $\chi^{0}$ as,
\begin{eqnarray}
\chi^{0}= 2\sqrt{kr_{c}}e^{-2kr_{c}\pi}\left( 1+\frac{3}{4}\omega^{2}e^{2kr_{c}\pi}\right) ^\frac{1}{2}\label{chi02}
\end{eqnarray}

The massless as well as massive modes thus depend on the induced brane
cosmological constant.\\

The masses for the various order KK modes as well as massless mode can be determined from zeros
of the Bessel function in the continuity condition at $\phi = \pi$. It is interesting to observe ( say from 
Equ.0.37) that
these values all are in the Tev range and do not change  significantly for a very wide range of  the values of the
cosmological constant $\omega^2$ say over a range of $1\le \omega^2 \ge 10^{-32}$. We present below the 
the masses of various modes of both KR and higher rank antisymmetric tensor field (see Table I).\\

\begin{table}[h]
\begin{tabular}{|c|cccc|}
\hline
$ n $ & 1 & 2 & 3 & 4 \\
\hline
$m_{n}^{tor}(for~KR~field)(TeV)$ & 3.75 & 7.015 & 10.173 & 13.323 \\
\hline
$m_{n}(for~higher~rank~tensor~field)(TeV)$ & 5.135 & 8.417 & 11.619 & 14.796 \\
\hline
\end{tabular}
\caption{Table of mass modes on anti de-Sitter 3-brane}
\end{table}
\section*{\large\bf Coupling with brane fermions}
Let us consider the interaction of both massless and the massive modes of the antisymmetric tensor fields 
with spin-$\frac{1}{2}$ fermions on the visible brane.
Starting from the 5-dimensional action and 
remembering that the fermion and its interactions are confined to the brane at $\phi=\pi$ we evaluate the coupling of the
KR field strength with the brane fermions. The fermion action is given as,
\begin{eqnarray}
S_{\psi}=i\int d^{4}x\int d\phi [detV]  \overline{\psi} [\gamma^{c} v^{\mu}_{a} (\partial_{\mu} - \frac{i}{2} G_{LN} \sigma^{ab} v^{\nu}_{a} \partial_{\mu} v^{\lambda}_{b} \delta_{\nu}^{N} \delta^{L}_{\lambda}-G_{AD}\sigma^{ab}v_{a}^{\beta}v_{b}^{\delta}\overline{\Gamma}^{A}_{MB}\delta_{\mu}^{M}\delta_{\beta}^{B}\delta_{\delta}^{D})]\psi \delta(\phi-\pi)\label{coupling1}\nonumber\\
\end{eqnarray}
where $G_{MN}$ is given by , 
\begin{eqnarray}
 G_{MN}=v^{a}_{M}v^{b}_{N}\eta_{ab}\label{coupling2}
\end{eqnarray}
and the vierbein $v_{\mu}^{a}$ is, 
\begin{eqnarray}
 v_{4}^{4}=1;v_{\mu}^{a}=e^{-A(\phi)}\delta_{\mu}^{a};detV=e^{-4A(\phi)}\label{coupling3}
\end{eqnarray}
a,b etc. being the tangent space indices.\\
Integrating out the compact dimension and using the fact that the fermion field on the brane is 
consistently renormalized as $\psi\rightarrow e^{3A(\pi)/2}\psi$, one obtains the effective 4-dimensional fermion  KR interaction as,
\begin{eqnarray}
&&\mathscr{L}_{\psi\overline{\psi}H}=i\overline{\psi}\gamma^{\mu}\sigma^{\nu\lambda}[ \frac{1}{M_{P}e^{kr_{c}\pi}}\left\lbrace 1+\frac{\omega^{2}}{4}\left( 1+e^{2 kr_{c}\pi}\right) \right\rbrace ^{\frac{1}{2}}H^{0}_{\mu\nu\lambda}+\nonumber\\ 
&&(1.18)\frac{m_{n}}{M_{P}k}\left( 1+\frac{1}{2}\frac{k^{2}\omega^{2}}{m_{n}^{2}}0.036\right) \left( J_{0}(x_{n})+\omega^{2}\xi^{n}\right)     \sum^{\infty}_{n=1}H^{n}_{\mu\nu\lambda}] \psi\label{coupling4}
\end{eqnarray}
where , $H^{n}_{\mu\nu\lambda}=\partial_{[\mu}B^{n}_{\nu\lambda]}$\\
Substituting the leading order approximation of $\xi^{n}$ from eq.(\ref{xin1}) in the eq.(\ref{coupling4})\\
We obtain,
\begin{eqnarray}
&&\mathscr{L}_{\psi\overline{\psi}H}=i\overline{\psi}\gamma^{\mu}\sigma^{\nu\lambda}[ \frac{1}{M_{P}e^{kr_{c}\pi}}\left\lbrace 1+\frac{\omega^{2}}{4}\left( 1+e^{2 kr_{c}\pi}\right) \right\rbrace ^{\frac{1}{2}}H^{0}_{\mu\nu\lambda}+\nonumber\\ 
&&(1.18)\frac{m_{n}}{M_{P}k}\left( 1+\frac{1}{2}\frac{k^{2}\omega^{2}}{m_{n}^{2}}0.036\right) \left( J_{0}(x_{n})-\omega^{2}\pi\frac{ke^{A\pi}}{m_{n}}J_{0}(x_{n})J_{1}(x_{n})Y_{0}(x_{n})\right)\nonumber\\ 
&&\sum^{\infty}_{n=1}H^{n}_{\mu\nu\lambda}] \psi\label{coupling4'}
\end{eqnarray}
The leading order coupling of massless KR field to the brane fermion now becomes
$\sim \frac{1}{M_{p}e^{-kr\pi}} + \frac{\omega}{M_{p}}$ 
whereas the leading order coupling of massive KR fields to the brane fermion is
$\sim \frac{e^{A(\pi)}}{M_{p}} + \frac{\omega^{2}e^{A(\pi)}}{M_{p}}$.\\
In the limit $\omega = 0$ we retrieve the expressions of the coupling as obtained in the flat brane scenario.
Though the corrections to the  couplings indeed depend on the brane cosmological constant but due  to the upper-bound 
( $\sim 10^{-32}$ )
on the magnitude of the induced brane cosmological constant in the anti- de-Sitter brane, 
both these corrections are vanishingly small on the visible brane. This implies that the
brane cosmological constant on the anti-de Sitter brane does not modify the result significantly from that in 
the flat 3-brane case and the massless mode again has extremely weak coupling whereas the massive modes have inverse Tev coupling. 
Thus the massless KR mode which can be identified with background space-time torsion still remains invisible 
in an anti-de Sitter warped geometry model.\\   

Next we take up the coupling of higher rank field strength tensor to the brane fermion located on the Ads brane.
Here we consider the rank-3 antisymmetric tensor field $X_{MNA}$ with the corresponding  
rank four field strength tensor $Y_{MNAB}$. Proceeding similarly as in case of the KR field, 
the final expression for the coupling becomes,
\begin{eqnarray}
&&\mathscr{L}_{\psi\overline{\psi}Y}=i\overline{\psi}\gamma^{\mu}\Sigma^{\nu\lambda\beta}[\frac{e^{A(\pi)}}{M_{p}e^{2kr_{c}\pi}}\left\lbrace 1+\frac{3}{4}\omega^{2}e^{2 kr_{c}\pi} \right\rbrace ^{\frac{1}{2}}Y^{0}_{\mu\nu\lambda\beta}+\nonumber\\
&&\frac{m_{n}}{M_{p}k}(2.71)(1-1.27\omega^{2})(J_{1}(x_{n})-\omega^{2}J_{1}(-\sqrt{1+a^{2}}x_{n}))\sum_{n=1}^{\infty}Y^{n}_{\mu\nu\lambda\beta}]\psi\label{coupling5}\nonumber\\
\end{eqnarray}
where , $a^{2}=\frac{k^{2}\omega^{2}}{m_{n}^{2}}$\\
It is evident fom the above expression that in absence of the cosmological constant i.e for a flat brane in RS scenario the
couplings for both the massless and massive modes are heavily suppressed.
The leading order correction to the coupling term for the massless higher rank tensor field
to the brane fermion is now $\sim \frac{e^{A(\pi)}\omega}{M_{p}e^{kr_{c}\pi}}$, 
while that for the massive higher rank tensor fields can be written as 
$\sim \frac{m_{n}\omega^{2}}{M_{p}k}$.
Both these corrections once again are very tiny due to the upper-bound of $10^{-32}$ on the value of the brane cosmological
constant in the anti de-Sitter case.\\
Thus we conclude that for anti de-Sitter brane their is not much change in the scenario of the presence for the
antisymmetric tensor fields on the anti de-Sitter 
brane from that in flat brane. All the massless and massive modes are heavily suppressed except
the massive modes for the rank two KR fields which has an inverse Tev coupling with the brane fermions.

We now shift our attention to the de-Sitter 3-brane solution in the  generalized RS model described earlier and examine 
the presence of various rank antisymmetric tensor fields on the visible brane. Our result reveals a drastic change of scenario
for the de-Sitter brane from that in an anti de-Sitter brane.

\section*{\Large\bf Modes of the Antisymmetric tensor fields in de-Sitter 3-brane}
Considering the induced brane cosmological constant on the visible 3-brane to be positive i.e $\Omega>0$  the warp 
factor in this case is , 
\begin{eqnarray}
 e^{-A(\phi)} = \omega  {\rm sinh}|({\rm ln}\frac{c_2}{\omega}-kr_{c}\phi)|\label{dswf}\\
 {\rm where} , c_{2} = 1+\sqrt{1+\omega^{2}} , \omega^{2}=\frac{\Omega}{3k^{2}}\nonumber
\end{eqnarray}
Unlike the Ads scenario, the induced cosmological 
constant in this case has no bound and the warp factor being different from the Ads scenario, the perturbed 
solution $\xi^{n}$ changes.
Repeating the same procedure as has been done for Ads brane , we can recast the  equation (\ref{dif1})
for $\chi^{n}$ in terms of $z_{n}=\frac{m_{n}}{k}e^{A(\phi)}$ as, 
\begin{eqnarray}
z_{n}^2\frac{d^{2}\chi^n}{dz_{n}^{2}}+z_{n}\frac{d\chi^{n}}{dz_{n}}-\frac{a^2z_{n}^{3}}{(1+a^{2}z_{n}^{2})}\frac{d\chi^{n}}{dz_{n}}+\frac{z_{n}^{2}\chi^{n}}{(1+a^{2}z_{n}^{2})}=0\label{znkrds1}
\end{eqnarray}
where, $a^{2}=\frac{k^{2}\omega ^{2}}{m_{n}^{2}}$
As the third term is small compared to $z_{n}$, we obtain
\begin{equation}
 [z_{n}^{2}\frac{d^{2}}{dz_{n}^{2}}+z_{n}\frac{d}{dz_{n}}+z_{n}^{2}(1-a^{2}z_{n}^{2})]\chi_{n}=0\label{znkrds2}
\end{equation}
The solution of the above equation can be written as,
\begin{equation}
\chi^{n}(\phi) = \frac{1}{N_{n}}[J_{0}(z_{n})+\alpha_{n}Y_{0}(z_{n})+\omega^{2}\xi^{n}]\label{chinkrds1}
\end{equation}
From continuity condition at $\phi=0$ we obtain , $\alpha_{n} \simeq x_{n}e^{-A(\pi)}$
where we have used $x_{n}=z_{n}(\pi)$.\\
This implies that  $\alpha_{n}<<1$ and therefore 
at $\phi=\pi$ we get, $J_{1}(x_{n})=\frac{\pi}{2}x_{n}e^{-A(\pi)}$\\

The differential equation for $\xi^{n}$ now becomes,
\begin{eqnarray}
\left[ z_{n}^{2}\frac{d^2}{dz_{n}^2}+z_{n}\frac{d}{dz_{n}}+z_{n}^{2}\right] \xi_{n}-\frac{k^{2}}{m_{n}^{2}}z_{n}^{4}J_{0}(z_{n})=0\label{znkrds3}
\end{eqnarray}
To the leading order $J_{0}(z_{n})=\frac{1}{2}$ the above differential equation becomes
\begin{eqnarray}
\left[ z_{n}^{2}\frac{d^2}{dz_{n}^2}+z_{n}\frac{d}{dz_{n}}+z_{n}^{2}\right] \xi_{n}-\frac{1}{2}\frac{k^{2}}{m_{n}^{2}}z_{n}^{4}=0\label{znkrds4}
\end{eqnarray}
The solution for $\xi^{n}$ is given as,
\begin{eqnarray}
\xi^{n}&=&\frac{1}{4}\frac{k^{2}}{m_{n}^{2}}[-2\pi z_{n}^2J_{0}(z_{n})Y_{0}(z_{n})+4\pi z_{n}J_{0}(z_{n})J_{1}(z_{n})Y_{0}(z_{n})+2\pi\nonumber\\
&&z_{n}^{2}J_{2}(z_{n})Y_{0}(z_{n})-\pi z_{n}^{3}J_{3}(z_{n})Y_{0}(z_{n})+4\pi z_{n}J_{0}(z_{n})Y_{1}(z_{n})\nonumber\\
&&-\pi z_{n}^{3}J_{0}(z_{n})Y_{1}(z_{n})-4\pi z_{n}J_{0}(z_{n})^{2}Y_{1}(z_{n})]\label{xinkrds1}
\end{eqnarray}
Using the  orthonormality condition (\ref{oc1}) and doing the  numerical integration we finally arrive at the  expression for $N_{n}$ as, 
\begin{eqnarray}
N_{n}=\frac{k}{m_{n}}\frac{1}{\sqrt{kr_{c}}}\left[0.714+\frac{k^{2}\omega^{2}}{m_{n}^{2}}0.0258 \right]^{\frac{1}{2}}\label{Nnkrds}
\end{eqnarray}
The final solution for the massive modes turns out to be, 
\begin{eqnarray}
\chi_{n}(z_{n})=\sqrt{kr_{c}}\frac{m_{n}}{k}\left[ 0.714+\frac{k^{2}\omega^{2}}{m_{n}^{2}}0.0258\right]^{-\frac{1}{2}} [J_{0}(z_{n})+\omega^{2}\xi^{n}]\label{chinkrds2}
\end{eqnarray}
To examine the presence of the massless mode of the antisymmetric tensor field on the visible brane , we turn our attention to  the 
differential equation for the massless mode which is now given as,
\begin{eqnarray}
 \frac{1}{r_{c}^{2}}\frac{d^{2}\chi_{n}}{d\phi^{2}}=0\label{mlmdifkrds}
\end{eqnarray}
The solution for the massless mode is obtained as,
\begin{eqnarray}
\chi^{0}(\phi)= c_{1}\phi+c_{2}
\end{eqnarray}
Here applying continuity condition once again we get, $c_{1}=0$. Use of the  orthonormality condition yields ,
\begin{eqnarray}
c_{2}^{2}=|2kr e^{-2kr_{c}\pi}[1-\frac{\omega^{2}}{4}(1+e^{2kr_{c}\pi})]|\label{c2krds}
 \end{eqnarray}
Putting the solution of $c_{2}$ in the final expression for $\chi^{0}$, we find,
\begin{eqnarray}
\chi^{0}=\sqrt{2}\sqrt{kr_{c}}e^{-kr_{c}\pi}|\left[ 1-\frac{\omega^{2}}{4}(1+e^{2kr_{c}\pi})\right] ^{\frac{1}{2}}|\label{mlmkrds}
\end{eqnarray}
This is the solution for the massless KR field on a de-Sitter 3-brane.

Let us now turn our attention to the bulk antisymmetric tensor field of  higher rank i.e a rank-3 tensor, $X_{MNA}$, 
with the corresponding field strength tensor $Y_{MNAB}$.
Following the procedure described so far and introducing $f_{n}=e^{A}(\phi)\chi^{n}$,
the equation (\ref{difyads}) in terms of $z_{n}=\frac{m_{n}}{k}e^{A(\phi)}$ can be recast as, 
\begin{eqnarray}
\left[ z_{n}^{2}\frac{d^{2}f_{n}}{dz_{n}^{2}}+z_{n}\frac{df_{n}}{dz_{n}}+\frac{f_{n}z_{n}^{2}}{(1+a^{2}z_{n}^{2})}-f_{n}-f_{n}\frac{a^{2}z_{n}^{2}}{(1+a^{2}z_{n}^{2})}\right] = 0\label{znyds1}
\end{eqnarray}
where, $a^{2}=\frac{k^{2}\omega^{2}}{m_{n}^{2}}$\\
Now we can reduce the equation (\ref{znyds1}) in terms of the perturbed solution $\xi^{n}$
\begin{eqnarray}
 z_{n}^{2} \frac{d^{2}\xi^{n}}{dz_{n}^{2}} + z_{n} \frac{d\xi^{n}}{dz_{n}} + [z_{n}^{2}(1-a^{2})-1]\xi^{n}=0\label{znyds2}
\end{eqnarray}
The solution for $\xi^{n}$ turns out to be, 
\begin{eqnarray}
\xi^{n}&=&\alpha_{n}Y_{1}(\sqrt{1-a^{2}}z_{n})-J_{1}(-\sqrt{1-a^{2}}z_{n})\nonumber\\
\chi^{n} &=& \frac{e^{-A(\phi)}}{N_{n}}[J_{1}(z_{n})+\alpha_{n}Y_{1}(z_{n})+\omega^{2}\{\alpha_{n}Y_{1}(\sqrt{1-a^{2}}z_{n})\nonumber\\
&&-J_{1}(-\sqrt{1-a^{2}}z_{n})\}]\label{chinyds1}
\end{eqnarray}
Applying the continuity 
condition at $\phi=0$ we again arrive at  $\alpha_{n}<<1$. This leads to,
\begin{eqnarray}
 \chi^{n}= \frac{e^{-A(\phi)}}{N_{n}}\left[ J_{1}(z_{n})-\omega^{2}J_{1}(-\sqrt{1-a^{2}}z_{n})\right]\label{chinyds2}
\end{eqnarray}
The continuity condition at $\phi=\pi$ yields,
\begin{eqnarray}
J_{2}(x_{n})+\sqrt{1-a^{2}}\omega^{2}J_{2}(\sqrt{1-a^{2}}x_{n}) = 0\label{besselyds}
\end{eqnarray}
Using equation (\ref{ocyads}) i.e the orthonormality condition and performing the numerical integration we get the expression for $N_{n}$,
\begin{eqnarray}
&&N_{n}=(0.368)\frac{k}{m_{n}}\frac{1}{\sqrt{kr_{c}}}\\
&&\chi^{n}=\frac{e^{-A(\phi)}m_{n}}{k}\sqrt{kr_{c}}(2.71)\left[ J_{1}(z_{n})-J_{1}(-\sqrt{1-a^{2}}z_{n})\}\right]\label{chinyds3}
\end{eqnarray}
This is the solution for the massive modes on the de-Sitter brane.\\
For the massless mode the solution is,
\begin{eqnarray}
\chi^{0}= c_{1}\left[ \frac{\omega^{4}e^{2kr_{c}\phi}}{32kr_{c}}-\frac{e^{-2kr_{c}\phi}}{2kr_{c}}-\frac{\omega^{2}\phi}{2}\right] + c_{2}\label{chi0yds1}
\end{eqnarray}
Applying continuity condition we again find,
$c_{1}=0$  and  $\chi^{0}=c_{2}$\\
Furthermore the orthonormality condition yields,
\begin{eqnarray}
 c_{2}^{2}=|4e^{-4kr_{c}\pi}[1-\frac{2}{3}\omega^{2}e^{2kr_{c}\pi}]|\label{c2yds}
\end{eqnarray}
Putting the solution of $c_{2}$,we get the final expression for $\chi^{0}$
\begin{eqnarray}
\chi^{0}= 2\sqrt{kr_{c}}e^{-2kr_{c}\pi}|\left( 1-\frac{2}{3}\omega^{2}e^{2kr_{c}\pi}\right) ^\frac{1}{2}|\label{chi0yds2}
\end{eqnarray}

Just as in anti de-Sitter case here also we determine various masses from the continuity condition at $\phi = \pi$ and
estimating the zeros of the Besel function. These are depicted in Table II. 
\begin{table}[htb]
\begin{tabular}{|c|cccc|}
\hline
$ n $ & 1 & 2 & 3 & 4 \\
\hline
$m_{n}^{tor}(for~KR~field)(TeV)$ & 3.726 & 6.996 & 10.17 & 13.27 \\
\hline
$m_{n}(for~higher~rank~tensor~field)(TeV)$ & 5.106 & 8.418 & 11.55 & 14.79 \\
\hline
\end{tabular}
\caption{Table of mass modes on de-Sitter 3-brane}
\end{table}
\section*{\large\bf Coupling with brane fermions}
We now consider the coupling of torsion as well as the higher rank antisymmetric tensor field to the matter fields on the 
de-Sitter visible brane. 

The warp factor for the de-Sitter case is 
\begin{eqnarray}
 e^{-A(\phi)} = \omega  {\rm sinh}({\rm ln}\frac{c_2}{\omega}-kr_{c}\phi)\nonumber\\
 {\rm where} , c_{2} = 1+\sqrt{1+\omega^{2}}\nonumber
\end{eqnarray}
Performing the same calculation as in the case of Ads brane for KR field , we get the 
expression for the coupling of torsion to the fermion ,residing on the visible brane as,
\begin{eqnarray}
&&\mathscr{L}_{\psi\overline{\psi}H}=i\overline{\psi}\gamma^{\mu}\sigma^{\nu\lambda}[ \frac{1}{M_{P}e^{kr_{c}\pi}}\left\lbrace 1-\frac{\omega^{2}}{4}\left( 1+e^{2 kr_{c}\pi}\right) \right\rbrace ^{\frac{1}{2}}H^{0}_{\mu\nu\lambda}+\nonumber\\ 
&&\frac{m_{n}}{M_{P}k}\frac{m_{n}}{\omega k}6.22 \left( J_{0}(x_{n})+\omega^{2}\xi^{n}\right)  
  \sum^{\infty}_{n=1}H^{n}_{\mu\nu\lambda}] \psi\label{coupling6}
\end{eqnarray}
Substituting the leading order approximation of $\xi^{n}$ from the eq.(\ref{xinkrds1}) in the eq.(\ref{coupling6})
\begin{eqnarray}
&&\mathscr{L}_{\psi\overline{\psi}H}=i\overline{\psi}\gamma^{\mu}\sigma^{\nu\lambda}[ \frac{1}{M_{P}e^{kr_{c}\pi}}\left\lbrace 1-\frac{\omega^{2}}{4}\left( 1+e^{2 kr_{c}\pi}\right) \right\rbrace ^{\frac{1}{2}}H^{0}_{\mu\nu\lambda}+\nonumber\\ 
&&\frac{m_{n}^{2}}{M_{P}k^{2}\omega}6.22 \left( J_{0}(x_{n})+\omega^{2}\pi\frac{ke^{A\pi}}{m_{n}}J_{0}(x_{n})J_{1}(x_{n})Y_{0}(x_{n}\right)  
  \sum^{\infty}_{n=1}H^{n}_{\mu\nu\lambda}] \psi\label{coupling6'}
\end{eqnarray}

In this case the leading order coupling of massless KR field to the brane fermion is
$\sim \frac{1}{M_{p}e^{kr_{c}\pi}}$ and  the leading order coupling of massive KR field to the brane fermion can 
be written as $\sim \frac{m_{n}\omega}{M_{p}k}e^{A(\pi)}$.\\
It may be observed from fig.(1) that the value of $\omega^2$ rises very steeply with the decrease in the value of $kr$ from the
corresponding RS value. It is given by the relation \cite{Ssg3}
\begin{eqnarray}
e^{-kr\pi} = \frac{10^{-16}[1 + \sqrt{1 + \omega^2 10^{32}}]}{ (1 + \sqrt{1 + \omega^2 })} 
\end{eqnarray}  
Due to the decrease in the value of $kr$ with increase in the value of the induced positive brane cosmological constant 
there will be a region where both the above couplings ( for massless and massive modes with brane fermions ) become
strong and can be comparable or larger than that of the gravity mode with the brane fermions.\\

Proceeding similarly the coupling of the higher rank tensor field with the brane localized fermion can be determined.
From the coupling term,
\begin{eqnarray}
&&\mathscr{L}_{\psi\overline{\psi}Y}=i\overline{\psi}\gamma^{\mu}\Sigma^{\nu\lambda\beta}[\frac{e^{A(\pi)}}{M_{p}e^{2kr_{c}\pi}}\left\lbrace 1-\frac{2}{3}\omega^{2}e^{2 kr_{c}\pi} \right\rbrace ^{\frac{1}{2}}Y^{0}_{\mu\nu\lambda\beta}+\nonumber\\
&&(2.71)\frac{m_{n}}{M_{p}k}(J_{1}(x_{n})-\omega^{2}J_{1}(-\sqrt{1-a^{2}}x_{n}))\sum_{n=1}^{\infty}Y^{n}_{\mu\nu\lambda\beta}]\psi\label{coupling7}
\end{eqnarray}
where , $a^{2}=\frac{k^{2}\omega^{2}}{m_{n}^{2}}$ \\
we can easily derive the leading order correction to the coupling term for the massless higher rank tensor field
to the brane fermion as $\sim \frac{e^{A(\pi)}}{M_{p}e^{2kr_{c}\pi}}$ \\
For large $\omega^2$, the exponential factor in the denominator can be small ( due to decrease in the value of $kr$ 
where as that in the numerator is
$\sim 10^{16}$. This leads to enhanced coupling for the massless modes.\\
The leading order coupling term for the massive higher rank tensor fields however can be written as, 
$\sim \frac{m_{n}\omega^{2}}{M_{p}k}$, which is suppressed by an additional factor of $k$ in the denominator and therefore
is heavily suppressed.
\section*{Conclusions}
Bulk antisymmetric tensor fields of two and higher rank have been studied in a generalized Randall-Sundrum model with
an induced cosmological constant on the visible 3-brane. We have considered both de-Sitter and anti de-Sitter non flat
3-branes with an appropriate warp factor which can resolve the gauge hierarchy problem in connection with the Higgs mass.
The massless modes of the bulk antisymmetric fields 
which have vanishingly small coupling with fermion matter field on the visible brane
in an usual RS scenario now acquires much larger coupling due to the presence of non-vanishing cosmological constant on
the 3-brane. It is shown that due to the constraints on the magnitude of the cosmological constant in an anti de-Sitter
3-brane in the generalized warped model these couplings continue to be small. However for de-Sitter 3-brane the decrease in
the value of modulus $kr$ alongwith a rise in brane cosmological constant enable to have a significantly large coupling
so that these antisymmetric tensor fields and their KK modes may have non-trivial role in particle phenomenology.
Such situation may be important in very early stage of the universe where a model with a large cosmological constant
is invoked to explain inflationary phase of the universe. Thus the antisymmetric tensor fields which are invisible
in the present epoch will be an inseparable part in describing Physics at the fundamental scale.

\end{document}